\documentclass[aps,prl,groupedaddress,twocolumn,showpacs,preprintnumbers,amsmath,amssymb]{revtex4-1}
\usepackage{graphicx}% Include figure files
\usepackage{dcolumn}% Align table columns on decimal point
\usepackage{bm}% bold math
\usepackage{upgreek}

\begin{document}

\title{ High bias anomaly in YBa$_2$Cu$_3$O$_{7}$/LaMnO$_{3+\delta}$/YBa$_2$Cu$_3$O$_{7}$
Superconductor/Ferromagnetic Insulator/Superconductor junctions:
Evidence for a long-range superconducting proximity effect
through the conduction band of a ferromagnetic insulator. }

\author{T. Golod}
\author{A. Rydh}
\author{V. M. Krasnov}
\email{Vladimir.Krasnov@fysik.su.se}
\affiliation{Department of
Physics, Stockholm University, AlbaNova University Center,
SE-10691 Stockholm, Sweden}

\author{I. Marozau}
\author{M.A. Uribe-Laverde}
\author{D.K. Satapathy}
\author{Th. Wagner}
\author{C. Bernhard}
\affiliation{University of Fribourg, Department of Physics and
Fribourg Centre for Nanomaterials, Chemin du Mus\'{e}e 3, CH-1700
Fribourg, Switzerland}

\date{\today }

\begin{abstract}
We study the perpendicular transport characteristics of small
superconductor/ ferromagnetic insulator/ superconductor
(YBa$_2$Cu$_3$O$_{7-x}$/ LaMnO$_{3+\delta}$/
YBa$_2$Cu$_3$O$_{7-x}$) tunnel junctions. At a large bias voltage
$V\sim 1$ V we observe a step-like onset of excess current that
occurs below the superconducting transition temperature $T<T_c$
and is easily suppressed by a magnetic field. The phenomenon is
attributed to a novel type of the superconducting proximity effect
of non-equilibrium electrons injected into the conduction band of
the ferromagnetic insulator via a Fowler-Nordheim tunneling
process. The occurrence of a strongly non-equilibrium population
is confirmed by the detection of photon emission at large bias
voltage. Since the conduction band in our ferromagnetic insulator
is strongly spin polarized, the long-range (20 nm) of the observed
proximity effect provides evidence for an unconventional
spin-triplet superconducting state.

\pacs{
75.47.Lx %Magnetic oxides
74.45.+c %Proximity effects; Andreev reflection; SN and SNS junctions
74.50.+r %Tunneling phenomena; Josephson effects (for SQUIDs, see 85.25.Dq; for Josephson devices, see 85.25.Cp; for Josephson junction arrays, see 74.81.Fa)
74.40.Gh %Nonequilibrium superconductivity
%74.78.Fk %superconducting Multilayers, superlattices,heterostructures
%75.76.+j %Spin transport effects (for devices exploiting spin polarized transport, see 85.75.Hh, 85.75.Mm, and 85.75.Ss)
%74.72.Gh %Hole-doped cuprates
%74.55.+v %Tunneling phenomena: single particle tunneling and STM
%85.75.Mm %Spin polarized resonant tunnel junctions
}
\end{abstract}

\maketitle

\section{I. Introduction}

The manganite perovskite La$_{2-x}$(Ca,Sr)$_{x}$MnO$_{3}$ exhibits a rich phase diagram of unusual electronic and magnetic properties. Most prominent is the so-called colossal magneto-resistance (CMR) effect which occurs at $0.15<x<0.5$ in the context of a transition from a paramagnetic and insulating or poorly conducting state to a ferromagnetic and half metallic state \cite{ManganitesReview,LMO_PLD}.
At somewhat lower doping at $0.1<x<0.15$, the corresponding transition leads into a ferromagnetic insulating (FI) state which is a
consequence of charge localization below the spin-polaron mobility
edge and orbital ordering at low $T$
\cite{Jahn_Teller_MillisPRB1996,MagPolaron_PRB1999,MagPolaron_NJP2004,STM_FischerPRB2008}. This FI state can also be achieved in off stoichiometric LaMnO$_{3+\delta}$ in which the hole doping is due to La or Mn vacancies.  While half-metallic manganites
are considered as promising spin-polarizers
\cite{ResTunnNanotube2}, the FI manganite can be an ideal
spin-filter material for use in tunnel barriers with low leakage
currents. Efficient spin-filters are on demand for spintronic
applications \cite{Moodera2011}. Spin polarization depends on many
parameters including the density of states (DOS) of electronic
bands
\cite{ResTunn,ResTunnNanotube1,ResTunnNanotube2,Niizeki,Zhang},
orbital hybridization and various material issues at interfaces
and on the bias voltage \cite{Fert2007}. Metallic ferromagnets
provide only modest spin polarization because both majority
($\uparrow$) and minority ($\downarrow$) spin bands are crossing
the Fermi level $E_F$ \cite{Meservey}. This problem is obviated in
insulating and half-metallic ferromagnets with separated majority and minority
bands at $E_F$, which, therefore, can facilitate an almost
complete spin-polarization
\cite{Moodera2011,Fert2007,Meservey,Blamire}.

The antagonism between superconductivity (S) and ferromagnetism (F) leads to a variety of novel effects in hybrid S/F heterostructures
\cite{Buzdin,Efetov2005}, such as the stabilization of unconventional spin-triplet superconductivity
\cite{Efetov2005,Golubov2007}, or the realization of $\pi$-Josephson
junctions \cite{Ryazanov}, that are interesting both from fundamental and
applied points of view. Electrons in a material, in contact with a
superconductor, inherit phase coherence from the S-side, a phenomenon known as the superconducting proximity effect. In clean
normal metals (N) the proximity effect can extend over a length scale
on the order of
micrometers. %\xi_N \gtrsim \mu$m.
In ferromagnetic metals the range depends on the superconducting
pairing symmetry \cite{Efetov2005,Golubov2007}: the conventional
spin-singlet pairing is rapidly suppressed by the magnetic
exchange field at the scale of $\sim 1$ nm \cite{Buzdin}, but the
unconventional spin-triplet pairing is immune to the magnetic
exchange field and the range is similar as in non-magnetic metals. Therefore long-range superconducting proximity effects, observed in
ferromagnetic metals \cite{Sosnin2007,Blamire2010,Khaire2010} and
half-metals
\cite{Keizer2006,Hu_LMO_Triplet2009,Dybko_LMO_Triplet2009,Kalcheim2011,Visani2012}
are taken as evidence for a spin-triplet superconducting state
induced via spin-active S/F interfaces
\cite{Efetov2005,Golubov2007}.

S/FI/S tunnel junctions are characterized by a non-trivial phase
dynamics \cite{Asano,Fogelstrom2000,Efetov2005} that is useful for
phase-coherent quantum spintronic devices \cite{Asano,Blamire}.
Atomically thin FI layers enable sizeable and almost fully
spin-polarized tunnel currents that are beneficial for spintronic
applications \cite{Moodera2011,Fert2007}. However, the
superconducting wave function and the tunneling probability decay
exponentially at the atomic scale with increasing FI thickness
\cite{Asano}. Thus, nm-thick insulators are impenetrable for
electrons due to the bandgap at the Fermi level $E_F$. Therefore,
generally the superconducting proximity effect is not expected in
insulators and particularly not in ferromagnetic insulators.

The situation may change in the presence of strong electric fields
or under non-equilibrium conditions. For example, it is well known
that a field $>10^7$ V/m can change the doping state of
semiconductors via ionization, or breaking of electronic bonds,
leading respectively to an avalanche, or Zener breakdown in p-n
junctions. A variety of electric field effects has been reported
in polarizable complex oxides \cite{Ahn_Oxides,Ueno2008},
including cuprates \cite{Koval,Motzkau} and manganites
\cite{Asamitsu,Rao,Srivastava2000,Wahl2003,Ma_2004,Hu_2005,Murakami,Moshnyaga,Mondal}.
In polar media the electric field can lead not only to charge
transfer but also to structural changes
\cite{Mondal,Murakami,Moshnyaga} and electromigration, which cause
resistive switching phenomena \cite{Waser2007}. In FI manganites
the electric field may depin immobile polarons, leading to
a non-linear electrical response at large bias voltage
\cite{Rao,Wahl2003}. The non-linear response at large bias is also
inherent for tunnel junctions with a thick insulating barrier.
The electric field leads to a linear distortion of the tunnel barrier
and facilitates Fowler-Nordheim type tunneling (field emission)
\cite{FN_theory} of hot electrons into the conduction band of the
insulator. Such non-equilibrium electrons in the conduction band
are mobile and in case of a FI are fully spin-polarized, which may
lead to a significant change of magnetoresistance (MR) effects
\cite{Nagahama2007}.

In this work we study the bias dependence of the perpendicular
magneto-tunneling characteristics of epitaxially grown nano-scale
S/FI/S junctions made of the cuprate superconductor
YBa$_2$Cu$_3$O$_{7-x}$ (YBCO) and the low-doped manganite LaMnO$_{3+\delta}$ (LMO) that is a ferromagnetic insulator below $T_{\mathrm{Curie}}\simeq 150$ K. It is
observed that at low bias voltage the electron transport through
the junctions is blocked due to the presence of the insulating
band gap (polaronic mobility edge) and the lack of direct
tunneling through the 20 nm thick LMO layer. However, at larger
bias, $V>0.3$ V, corresponding to the regime of Fowler-Nordheim tunneling
into the conduction band of the insulator, excess current through
the junction appears. Initially it increases gradually with
increasing bias voltage but at $V\sim 1$ V it jumps abruptly,
leading to a step-like anomaly in the Current-Voltage
characteristics. Such step-like increase of the excess current exists only in the
superconducting state and is easily suppressed by a magnetic field.
The phenomenon is attributed to a novel type of the
superconducting proximity effect of non-equilibrium electrons in
the conduction band of the ferromagnetic insulator. The occurrence of this strongly
non-equilibrium state in the LMO layer is confirmed by the direct
detection of photon emission. Since the conduction band in our
ferromagnetic insulator is fully spin polarized, the long-range
(20 nm) of the observed proximity effect points towards an unconventional spin-triplet superconducting state.

The paper is organized as follows. In Sec. II we describe
the fabrication procedure of the YBCO/LMO heterostructures and the nano-scale
YBCO/LMO/YBCO tunnel junctions. We demonstrate the
reproducibility of the ferromagnetic and insulating properties of LMO layers both for a single
layer and within YBCO/LMO multilayers with different layer
thicknesses. In Sec. III we describe the perpendicular transport
properties of the LMO layer in our junctions. It is shown that it
undergoes a transition to the ferromagnetic insulator state at
$T_{\mathrm{Curie}}\simeq 150$ K with a strongly bias-dependent CMR up to
-600$\%$. In Sec. IV we report the observation of a high-bias
anomaly in the junction characteristics, we demonstrate the occurrence of
a Fowler-Nordheim tunneling into the conduction band of the FI and
we provide an interpretation of the anomaly in terms of a novel
type of superconducting proximity effect of non-equilibrium
electrons in the conduction band of the ferromagnetic insulator.
The presence of a strongly non-equilibrium electron and hole
population in the conduction and valence bands of the FI is
confirmed by the direct detection of photon emission from our
junctions at large bias. Finally we argue that the long range of
the proximity effect through the FI indicates the unusual
odd-spin-triplet nature of the superconducting order parameter. We
also argue that the Fowler-Nordheim tunneling in the FI should
lead to perfectly spin-polarized currents that are beneficial for
spintronic applications.

\section{II. Samples}

\begin{figure*}[t]
    \centering
    \includegraphics[width=\textwidth]{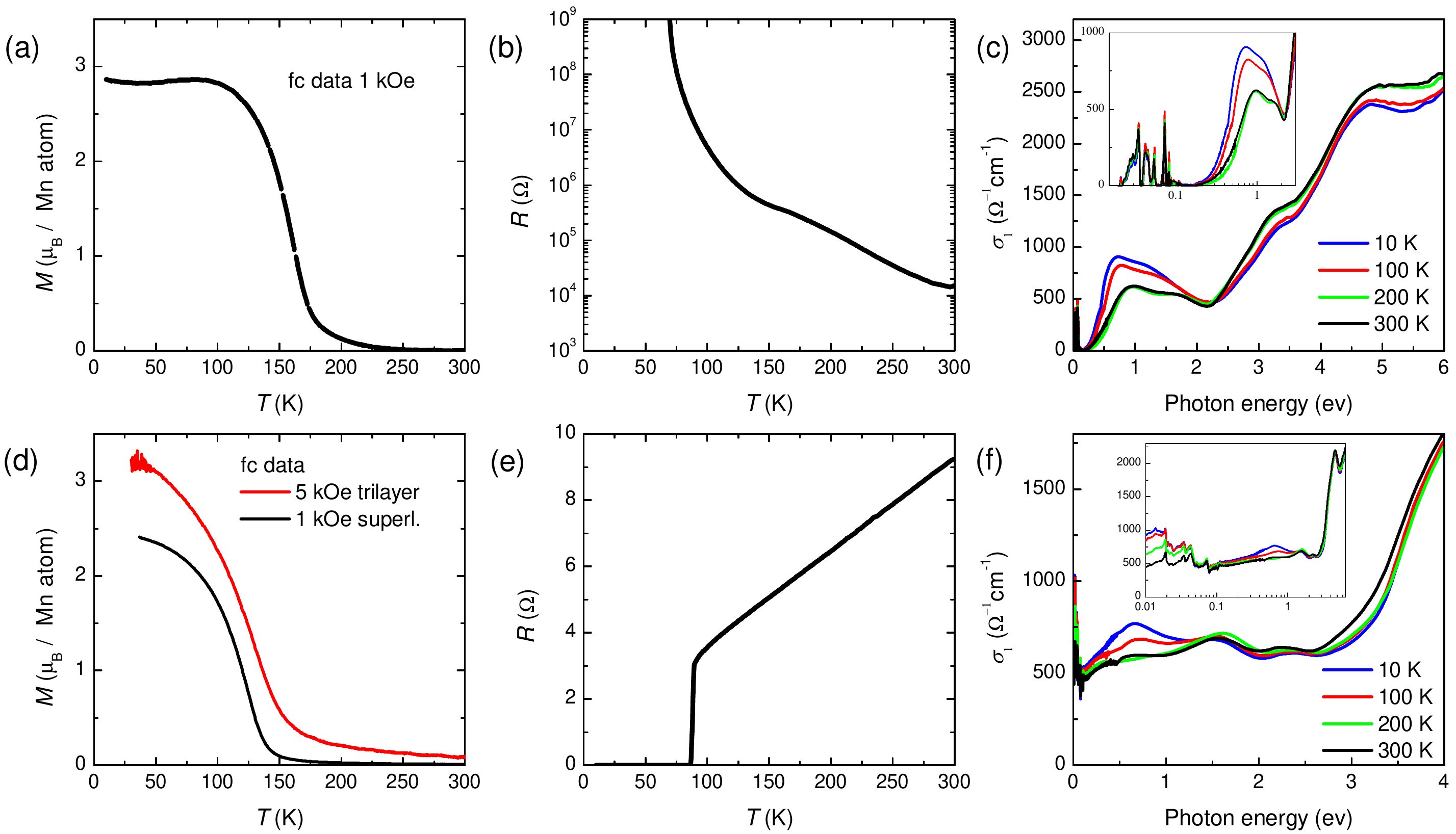}
    \caption{(Color online).
Magnetisation (a), resistance (b) and optical spectra (c) of a
single 100 nm thick LMO film showing its ferromagnetic insulator
properties. (d) Field-cooled magnetisation of a YBCO (10nm)/LMO
(10nm)/YBCO (10nm) trilayer at 5 kOe and a [YBCO (10nm)/LMO
(10nm)]x10 superlattice at 1 kOe. (e) Lateral resistance of the
YBCO (10nm)/LMO (10nm)/YBCO (10nm) trilayer with the contacts on
the top YBCO layer. (f) Optical conductivity of the [YBCO
(10nm)/LMO (10nm)]x10 superlattice.}
    \label{fig:fig1_LMO}
\end{figure*}

\subsection{A. Fabrication and characterization of YBCO/LMO heterostructures}

The YBCO/LMO heterostructures were deposited by pulsed laser
deposition (PLD) on (001)-oriented
Sr$_{0.7}$La$_{0.3}$Al$_{0.65}$Ta$_{0.35}$O$_3$ (LSAT) substrates
similar as described in Refs. \cite{Malik}. The substrate
temperature was $900\,^{\circ}\mathrm{C}$, the O$_2$ partial pressure 0.32 mbar and
the fluency and pulse frequency of the KrF 248 nm excimer laser
were 2 J/cm$^{2}$ and 2 Hz, respectively. The growth dynamics was
monitored with in-situ reflection high energy electron diffraction
(RHEED) to ensure a 2-dimensional growth mode with sharp
interfaces between the individual layers. After the growth the
sample was cooled to $700\,^{\circ}\mathrm{C}$ at $10\,^{\circ}\mathrm{C}$/min where the oxygen pressure
was increased to 1 bar. Subsequently, the sample was further
cooled to $485\,^{\circ}\mathrm{C}$ at a rate of $30\,^{\circ}\mathrm{C}$/min where it was annealed for
one hour before it was cooled to room temperature. Finally, the structures were covered by a protective gold layer (70 nm).

For comparison, we also grew single LMO layers on LSAT substrates.
It is well known that the electronic and magnetic properties of
these LMO films can be largely varied depending on the PLD growth
conditions \cite{LMO_PLD}. We studied extensively how these can be
changed in a systematic way from an insulating and
antiferromagnetic or glassy magnetic state, over an insulating
ferromagnetic state to a metallic ferromagnetic state with a
modification of the O$_2$ background gas pressure during PLD
growth and a subsequent post-annealing treatment in flowing oxygen
or argon. This study, which will be presented in full length in a
forthcoming publication, has established that the above mentioned
growth conditions (an O$_2$ background pressure of 0.32 mbar and
post-annealing in flowing O$_2$ at $485\,^{\circ}\mathrm{C}$)
yield LMO layers that are insulating and strongly ferromagnetic
with a ferromagnetic moment of about 3 $\mu_B$ per Mn atom. The
post-annealing treatment does not strongly affect the FI
properties of the LMO layer but is required to fully oxygenate the
YBCO layers as to achieve a sharp superconducting transition with
a high critical temperature $T_c$.

The films were investigated with in-plane four-probe
electric transport and magnetization measurements with a Quantum Design physical property measurement
system (PPMS QD6000) with a VSM option. To measure the high
resistance of the single layer LMO films we used an external
Keithley multimeter. The ellipsometry measurements were performed
with a commercial rotating analyzer setup (Woollam VASE) in the
range 0.5-6.5 eV and with a homebuilt one at 0.01-0.5 eV
\cite{Bernhard2004}. The substrate correction was performed with
the commercial WVASE32 software package.

The FI properties of a 100 nm thick single LMO layer are
demonstrated in Figs.~\ref{fig:fig1_LMO}~(a-c). The field-cooled
magnetization data at 1 kOe in Fig.~\ref{fig:fig1_LMO}~(a) show
that a ferromagnetic state develops below $T_{\mathrm{Curie}} \simeq 160$ K
with a sizeable low temperature moment of about 2.8 $\mu_B$ per Mn ion. The
insulating behavior is evident from the temperature dependence of
the resistance shown in Fig.~\ref{fig:fig1_LMO}~(b) as well as
from the optical spectra in Fig.~\ref{fig:fig1_LMO}~(c). The
latter show a characteristic redistribution of spectral weight
below $T_{\mathrm{Curie}}$ from high energies above 2 eV towards two low
energy bands that are peaked around 1.3 eV and 0.7 eV,
respectively. A similar trend was previously reported for
La$_{1.9}$Sr$_{0.1}$MnO$_3$ in the ferromagnetic insulating state
\cite{Okimoto} where the low-energy band was assigned to the
response of strongly pinned polarons of orbital and/or magnetic
origin \cite{MagPolaron_NJP2004,MagPolaron_PRB1999}. The optical
spectra and the resistivity data confirm that these polarons are
strongly pinned and therefore do not contribute to a coherent
charge transport. As shown in the inset, the peaks at low
frequency are entirely due to infrared-active phonon modes. This
is different from the ferromagnetic, metallic state of
La$_{2/3}$(Ca,Sr)$_{1/3}$MnO$_3$ for which the spectral weight
transfer from high to low energy gives rise to a pronounced
Drude-like peak in the optical conductivity.

%\begin{figure*}[t]
%    \centering
%    \includegraphics[width=0.9\textwidth]{Fig2_HeterostrCharact.pdf}
%    \caption{(Color online).
%(a) Field-cooled magnetisation of a YBCO (10nm)/LMO (10nm)/YBCO
%(10nm) trilayer at 5 kOe and a [YBCO (10nm)/LMO (10nm)]x10
%superlattice at 1 kOe. (b) Lateral resistance of the YBCO
%(10nm)/LMO (10nm)/YBCO (10nm) trilayer with the contacts on the
%top YBCO layer. (c) Optical conductivity of the [YBCO (10nm)/LMO
%(10nm)]x10 superlattice. }
%    \label{fig:fig2_Heterostr}
%\end{figure*}

Figures~\ref{fig:fig1_LMO}~(d-f) demonstrate the reproducibility
of the characteristics of YBCO/LMO multilayers with different
number of thin (10 nm) layers. It is seen that the thin LMO layers
maintain their insulating ferromagnetic properties even if they
are combined with YBCO in the form of YBCO/LMO multilayers. Fig.~\ref{fig:fig1_LMO}~(d) shows the temperature dependent
magnetization of a YBCO(10nm) / LMO (10nm) /YBCO (10 nm) trilayer
and a [YBCO (10nm)/LMO (10nm)]x10 superlattice that were grown
under similar conditions. They both reveal a ferromagnetic
transition of the LMO layers at $T_{\mathrm{Curie}} \simeq 150$ K
with a sizeable magnetic moment of about 3 $\mu_B$ per Mn ion.
Fig.~\ref{fig:fig1_LMO}~(e) shows the corresponding $T$-dependence
of the in-plane resistance for the unpatterned YBCO/LMO/YBCO
trilayer. The in-plane resistance is dominated by the upper YBCO
layer on which the contacts are placed and does not exhibit any
noticeable contribution from the LMO layer that is situated
underneath. The resistance exhibits a linear $T$-dependence in the
normal state and a sharp superconducting transition around $T_c
\simeq 88$ K that is characteristic of the response of optimally
doped YBCO. However, direct evidence that the LMO layers maintain
their insulating properties in these YBCO/LMO multilayers has been
obtained from the optical conductivity of a corresponding [YBCO
(10nm) / LMO (10nm)]x10 superlattice as shown in Fig.~\ref{fig:fig1_LMO}~(f). The spectra also reveal the characteristic
spectral weight redistribution from high energy above 1.5 eV
towards a low energy peak around 0.65 eV that occurs right below
the ferromagnetic transition of the LMO layers at
$T_{\mathrm{Curie}} \simeq 150$ K. Similar to the single LMO
layer, for which the spectra are shown in Fig.~\ref{fig:fig1_LMO}~(c), this low-energy peak can be assigned to the strongly pinned
magneto-polarons. The additional Drude response which is present
at all temperatures originates from the metallic YBCO layers.

\begin{figure*}[t]
    \centering
    \includegraphics[width=\textwidth]{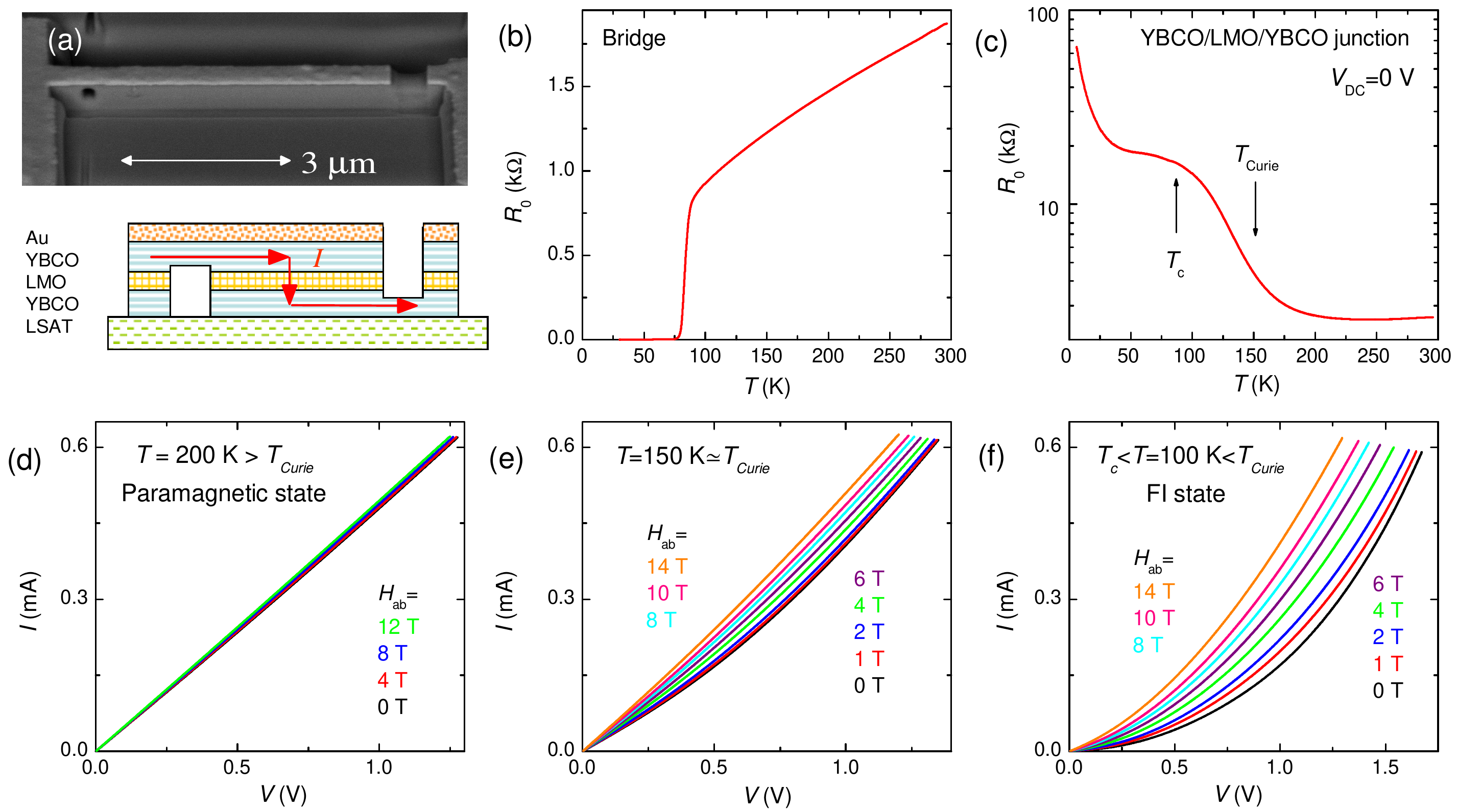}
    \caption{(Color online).
(a) Top panel: SEM image of a 3D-FIB patterned junction $\#2$.
%The junction width is $\sim 660$ nm.
Bottom panel: side-view sketch of the structure (not in scale).
Arrows indicate the current flow path. (b) In-plane resistive
transition of the same bridge before the final FIB cut. A sharp
superconducting transition occurs at $T_c \simeq 86$ K. (c)
Temperature dependence of the out-of-plane zero-bias resistance
$R_0(T)$ (in the semi-logarithmic scale) of the junction $\#2$ at
$H=0$. The transition to the insulating behavior occurs at
$T<T_{\mathrm{Curie}}\simeq 150$ K. Note that $R_0(T)$ reveals no
sign of superconductivity at $T<T_c$, indicating the lack of a
proximity effect at low bias. (d-f) Magnetic field dependence of
the $I$-$V$ curves of the junction $\#4$: (d) in the paramagnetic
state $T=200$ K, (f) close to the Curie temperature
$T=150\,\textrm{K} \simeq T_{\mathrm{Curie}}$, and (f) in the FI
state $T=100\,\textrm{K} < T_{\mathrm{Curie}}$. Negative colossal
magnetoresistance appears in the FI state.}
    \label{fig:fig2}
\end{figure*}

\subsection{B. Fabrication of 3D-nanosculptured YBCO/LMO/YBCO junctions}

To study the perpendicular transport properties of the LMO layer, we
fabricated small S/FI/S junctions using a 3D-nanosculpturing
technique \cite{Golod_PRL2010}. The growth of the YBCO/LMO/YBCO (100/20/100 nm) heterostructure, covered by a protective Au
layer (70 nm), was described in the previous Sec. II A. This heterostructure was
patterned into several $\sim 5$ $\mu$m wide bridges using
photolithography and cryogenic (liquid nitrogen-cooled) reactive
ion etching in Ar-plasma. At the next stage the central part of
each bridge was narrowed to a sub-micron width using a focused ion
beam (FIB), as shown in Fig.~\ref{fig:fig2}~(a). Finally, two
side cuts were made by FIB to interrupt the bottom YBCO and the
top YBCO/Au layers. This leads to the appearance of a zig-zag
structure as sketched in the bottom part of Fig.~\ref{fig:fig2}~(a). The arrow indicates the current flow path. The side-cuts
force the current to flow through the LMO layer, thus a small
YBCO/LMO/YBCO (S/FI/S) tunnel junction is formed between the two
side-cuts. In total three working junctions with similar characteristics were fabricated and studied.

Fig.~\ref{fig:fig2}~(b) shows the zero dc-bias (measured with small
ac-bias current) in-plane resistive transition $R_0(T)$ of the $660$
nm wide bridge ($\#2$) in Fig.~\ref{fig:fig2}~(a), before
the final side-cut through the top YBCO layer, but with an
interrupted Au layer. A sharp superconducting transition at
$T_c \simeq 86$ K is evident. A comparison with the superconducting
transition for the unpatterned multilayer in Fig.~\ref{fig:fig1_LMO}~(e), indicates that no significant deterioration of the YBCO
layers has occurred during the patterning process. This is due to a low oxygen
out-diffusion during cryogenic Ar-plasma etching and a
sufficient thickness of the protective Au layer that prevents
chemical passivation during the lithographic process and Ga-implantation during FIB etching/viewing.

\section{III. Perpendicular transport properties of the LMO layer}

In the following we present the transport measurements after the two-side cuts were performed with the FIB to interrupt the bottom and top YBCO layers. Arrows in the bottom part of Fig.~\ref{fig:fig2}~(a) indicate the
current flow path in our junctions. The current is forced to flow
across the LMO layer (in the $c$-axis direction) within the
junction area. This allows direct probing of the perpendicular
transport characteristics of the LMO layer inside the junctions.

\subsection{A. Temperature dependence of the $c$-axis resistance of LMO}

Fig.~\ref{fig:fig2}~(c) displays the temperature dependence of the
zero bias resistance $R_0(T)$ (measured with a small ac-current)
at zero applied magnetic field $H=0$ for junction $\#2$, as shown
in Fig.~\ref{fig:fig2}~(a). A comparison with the data for the
same bridge before the final side-cut through the top YBCO layer,
Fig.~\ref{fig:fig2}~(b), reveals that the out-of-plane $R_0(T)$ of
the junction differs both quantitatively and qualitatively from
the in-plane $R_0(T)$ of the bridge. From room temperature to
about 220 K the junction resistance decreases slightly with
decreasing temperature.
%showing a metallic trend.
%Down to $T\gtrsim 150$ K $R_0(T)$ of the junction is almost constant.
However, at $T<150$ K the resistivity rapidly starts to increase with
decreasing temperature, i.e., it shows an insulator-type behavior. At $T
=100$ K the resistance of the junction is about 20 times larger
than the one of the bridge. Although a minor drop of $R_0$
%of few hundred Ohm
due to the superconducting transition of the YBCO electrodes is
present, as marked by the upward arrow in Fig.~\ref{fig:fig2}~(c), it
is hardly visible on the scale of the plot. This demonstrates that
the junction resistance at low temperatures is governed by the resistance of the LMO barrier. The $c$-axis
resistivity of LMO at $T_c<T<T_{\mathrm{Curie}}$ with $\rho_c\sim 200-250
~\Omega$cm is clearly non-metallic. Therefore, the rapid upturn of
$R_0(T)$ at $T<150$ K indicates a transition of LMO into the
insulating state. This temperature is similar to the Curie
temperature of our single LMO film and coincides with $T_{\mathrm{Curie}}$
of LMO within the YBCO/LMO heterostructures as shown in Fig.~\ref{fig:fig1_LMO}.
%, as marked by the downward arrow in Fig.~\ref{fig:fig2}~(c).
The resistance continues to rapidly increase at $T<50$ K, showing
a tendency for diverging at $T\rightarrow 0$, in a qualitative
similarity with the behavior of the in-plane resistance of the
single LMO film, shown in Fig.~\ref{fig:fig1_LMO}~(b).

\begin{figure*}[t]
    \centering
    \includegraphics[width=\textwidth]{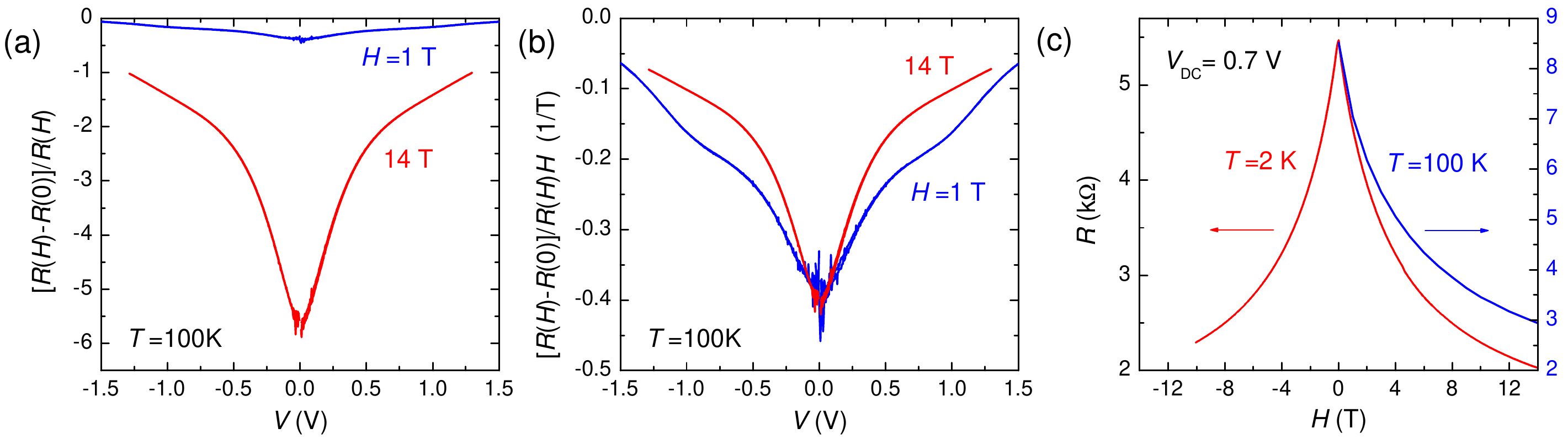}
    \caption{(Color online).
(a) Bias dependence of the magnetoresistance (MR) at in-plane
fields of $H=1$ and 14 T for the junction $\#4$. The MR reaches
-600$\%$ at $H=14$ T. (b) The same as in (a), normalized by the
magnetic field. It is seen that at low bias $|V|<0.1$ V the MR is
approximately linear in field up to 14 T, at intermediate bias the
MR is non-linear in $H$ and it again becomes approximately linear
at a higher bias $|V|\gtrsim 1.5$ V. (c) Non-linear
field-dependence of the junction resistance at an intermediate
bias $V=0.7$ V at $T=2$ (left axis) and 100 K (right axis). }
    \label{fig:fig3}
\end{figure*}

The comparison of Figs.~\ref{fig:fig1_LMO}~(d) and \ref{fig:fig2}
(c) demonstrates that the upturn of $R_0(T)$ at $T<150$ K follows
the one of the magnetization. As shown below, it is also
accompanied by the appearance of the colossal magnetoresistance,
which is typical for ferromagnetic manganites.
%reaches -600$\%$ at $H=14$ T. The CMR
%and is attributed to the double-exchange mechanism \cite{ManganitesReview}.
This proves that our LMO layer indeed enters the FI state below $T_{\mathrm{Curie}} \simeq 150$ K.

\subsection{B. Bias dependence of the colossal magnetoresistance}

The small area of our junctions allows achieving significant bias
voltages at small currents and without significant self-heating
\cite{Heating}. This facilitates an accurate analysis of the genuine
bias dependence of the perpendicular transport characteristics of LMO, without artifacts from self-heating.

Figs.~\ref{fig:fig2}~(d-f) show the current-voltage ($I$-$V$)
characteristics of the junction $\#4$ for different in-plane
magnetic fields: (d) in the paramagnetic state at $T=200\:\mathrm{K}>T_{\mathrm{Curie}}$, (e) close to the Curie temperature at $T=150$ K, and (f) in the FI state at $T=100\:\textrm{K} < T_{\mathrm{Curie}}$. It is seen that in the paramagnetic state the $I$-$V$'s are Ohmic and do not
exhibit a significant MR. However, already at 150 K and even more so at 100 K the $I$-$V$ curves become
non-linear and exhibit a large negative MR.

\begin{figure*}[t]
    \includegraphics[width=\textwidth]{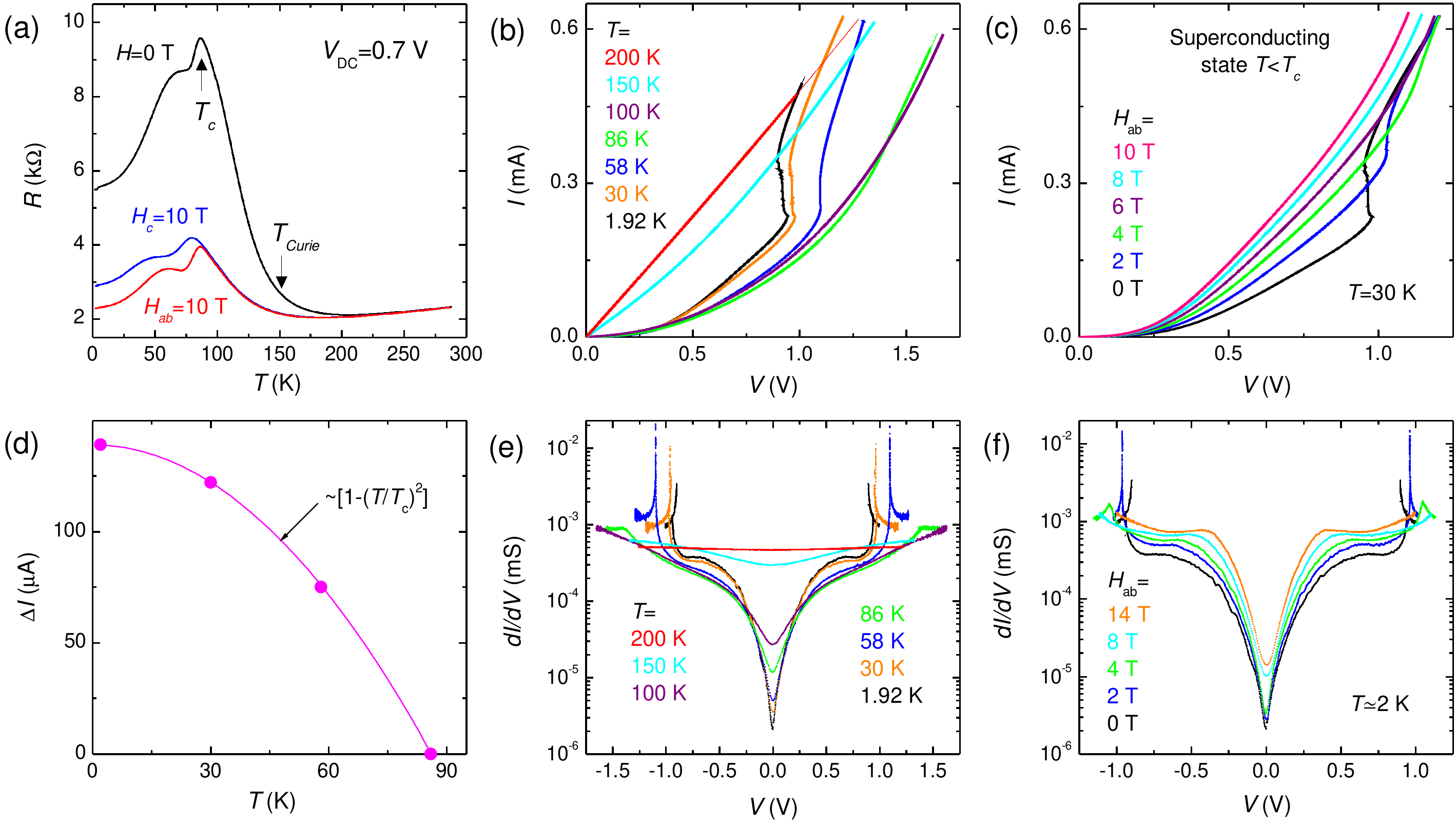}
    \caption{(Color online).
(a) Large bias %(above the polaronic mobility edge in LMO)
dc-resistance of the junction $\#4$ at $H=0$ and 10 T applied
along the $ab$-plane and the $c$-axis direction. The upturn at
$T<T_{\mathrm{Curie}}\sim 150$ K marks the transition to the FI
state of LMO while the downturn at $T_c \simeq 86$ K indicates the
appearance of a superconducting proximity effect. (b) $I$-$V$
curves of the junction $\#4$ at different $T$ and at $H=0$. Below
$T_c$, a step-like onset of excess current appears at large bias
voltage of about 1 V. (c) Magnetic field dependence of the $I$-$V$
curves at $T=30$ K. The step-like increase of the current is
rapidly suppressed by a magnetic field. (d) $T$-dependence of the
excess current (step height) $\Delta I$ at high bias. (e) and (f)
differential conductance $dI/dV$ of the junction $\#4$ (e) at
different temperatures and $H=0$, (f) at different $H$ and
$T\simeq 2$ K. }
    \label{fig:fig4}
\end{figure*}

Fig.~\ref{fig:fig3}~(a) shows the bias-dependence of the
magnetoresistance, $\textrm{MR}(V)=[R(V,H)-R(V,0)]/R(V,H) =
[1-I(V,H)/I(V,0)]$, for junction $\#4$ at $T=100$ K for in-plane
fields $H=1$ and 14 T. At low bias the $\textrm{MR}\,(V\sim 0)$
reaches -600$\%$ at $H=14$T, which is a clear signature of the CMR
effect. With increasing bias the CMR rapidly decreases and falls
to -100$\%$ at $V=1.25$ V. The bias dependence of the MR at
$V>0.1$ V approximately follows the power law $\textrm{MR}(V)
\propto -|V|^{-0.7}$. Panel (b) shows the same data normalized by
the magnetic field. It is seen that the field-normalized MR
coincides both at low $|V|<0.1$ V and high $|V|>1.5$ V bias, but
deviates at intermediate bias. This indicates that the MR at low
and high bias is linear in field at least up to $H=14$ T. However
the MR becomes significantly non-linear at intermediate bias of
$V\sim 1$ V. This is demonstrated in Fig.~\ref{fig:fig3}~(c),
which shows the field-dependence of the dc-resistance $R=V/I$ at
$V=0.7$ V. The MR is isotropic with respect to field orientation,
as shown in Fig.~\ref{fig:fig4}~(a).

The observed bias-dependence of the MR is similar to the one that was previously
reported for magnetic tunnel junctions containing LMO layers
\cite{Parkin1998,Fert2007,Hayakawa2002}. The strong
bias-dependence is still not well understood because it may be
caused by various intrinsic and extrinsic \cite{extrinsic} mechanisms. Intrinsic mechanisms for the suppression of the MR at high bias can
be due to inelastic spin-flip scattering of high-energy electrons
\cite{Tsui1971}, or to a strong energy dependence of the
electronic DOS near $E_F$ \cite{dos,Moodera2011}. Extrinsic
reasons can be due to granularity and microscopic inhomogeneity of
films \cite{Parkin1998,extrinsic}, or to oxygen variation at
interfaces \cite{Hayakawa2002}.

Since our junctions contain only one magnetic layer, they are not
of the spin-valve type. The epitaxial LMO layer is uniform in the
$c$-axis direction at the scale of its thickness 20 nm. Therefore,
it is unlikely that the observed strong bias-dependence of the MR
is of extrinsic origin, for example due to a spin-valve effect
that might occur in a granular structure. Rather it is an
intrinsic property of the LMO material. As described in Sec. IV C
below, the reduction of the MR at large bias is accompanied by a
significant photon emission. This shows the occurrence of
inelastic scattering in the LMO barrier and points towards an
intrinsic, spin-flip scattering mechanism \cite{Tsui1971} of the
strong bias dependence of the CMR in LMO.

\section{IV. Tunneling characteristics of YBCO/LMO/YBCO junctions in the superconducting state}

Unlike the MR of LMO, the electronic and superconducting
properties of the cuprates are strongly anisotropic. An
out-of-plane magnetic field suppresses superconductivity due to
the penetration of Abrikosov vortices
\cite{Fischer1995,MR_PRB2011}. However, an in-plane field of 10-17
T does not give rise to any visible suppression of
superconductivity at $T \ll T_c$ (see e.g. Fig.~3~(d) in Ref.
\cite{MR_PRB2011}). This is because the upper critical field
$H_{c2}$ of optimally doped cuprates is very high for the field
parallel to $ab$-planes. Therefore, to avoid complications with
field screening and suppression of superconductivity in YBCO, in
what follows we will focus on junction characteristics for
in-plane magnetic fields, that practically do not affect the YBCO
electrodes.

%15-20$ mV develop in $dI/dV$ at low $T$, which is accompanied by
%rapid suppression of the zero-bias conductance. The maxima are
%easily suppressed by the $c$-axis, Fig.~\ref{fig:fig4}~(b), but
%not the $ab$-plane magnetic field, Fig.~\ref{fig:fig4}~(c). Such
%behavior is typical for the sum-gap peak in superconducting tunnel
%junctions at $V=2\Delta/e$, where $\Delta$ is the superconducting
%energy gap \cite{MR_PRB2011}. The anisotropic response to magnetic
%field is inherent to YBCO films, which is contrast to the
%isotropic CMR in the LMO layer at $T>T_c$, see Fig.~\ref{fig:fig5}
%(a). Since the upper critical field $H_{c2}$ of optimally doped
%cuprates is very high for the field parallel to $ab$-planes, the
%in-plane field of 10-17 T does not do visible suppression of
%superconductivity at $T \ll T_c$ (see e.g. Fig.~3~(d) in Ref.
%\cite{MR_PRB2011}). From comparison of $dI/dV(V)$ curves at $H=0$
%in Fig.~\ref{fig:fig4}~(b) and in-plane $H=10$ T in Fig.~%\ref{fig:fig4}~(c) it is seen that the in-plane field of 10 T does
%not affect the superconducting sum-gap feature, unlike the
%out-of-plane field, that strongly suppresses the sum-gap maxima,
%as seen from Fig.~\ref{fig:fig4}~(b). Keeping in mind that the
%magnetoresistance in LMO is isotropic, to avoid complications with
%field screening and suppression of superconductivity in YBCO, in
%what follows we will focus on junction characteristics in the
%in-plane magnetic field, that practically does not affect YBCO
%electrodes.

Cooper pair tunneling in superconducting tunnel junctions should
lead to the appearance of a Josephson supercurrent at $V=0$.
%and a sum-gap peak weak in $dI/dV$ at maxima at $V=\pm
%2\Delta/e\simeq 50-80$ mV, where $\Delta$ is the superconducting
%energy gap \cite{Fischer1995,MR_PRB2011}.
However, as seen from Fig.~\ref{fig:fig2}~(c), the zero bias
resistance of the junction does not drop to zero at $T<T_c$.
Instead, it continues to rapidly increase with decreasing $T$.
Thus, there is no sign of a Josephson current at low bias. Our
FI layers are too thick (20nm) for the Cooper pair tunneling that is needed
for achieving the Josephson coupling. Since the probability of
Cooper pair and direct quasiparticle tunneling are the same,
direct quasiparticle tunneling through the LMO layer in our
junctions is also negligible.

\subsection{A. High-bias anomaly: evidence for
the superconducting proximity effect in LMO}

Fig.~\ref{fig:fig4}~(a) shows the temperature dependence of the
dc-resistance at an intermediate bias of $V=0.7$ V for the junction
$\#4$ at $H=0$ and $10$ T parallel ($ab$-plane) and
perpendicular ($c$-axis) to the film surface. It is seen that at $T>T_c$
the MR is isotropic with respect to the field orientation, which
is typical for manganites. An anisotropy below $T_c$ is caused by
a partial screening of the perpendicular field by the superconducting
YBCO layers. The comparison of Figs.~\ref{fig:fig2}~(c) and
\ref{fig:fig4}~(a) reveals a remarkable, qualitative difference in the
temperature dependence of the junction resistances at low and high
bias voltages in the superconducting state. At zero bias $R_0(T)$ grows
monotonously with decreasing $T$, showing no sign of a Josephson
coupling or superconducting proximity effect through the LMO
layer. As was discussed above, this is expected since the LMO layer
is insulating and too thick for direct tunneling. The absence of a proximity
effect at low bias voltage proves that our junctions are free from
microshots or extended non-magnetic ``dead" layers \cite{Ryazanov}
at the S/F interfaces, in which a long-range superconducting
proximity effect might be induced. This suggests that the very different behavior of $R_{\mathrm{dc}}$ at higher bias voltages is an intrinsic effect due to the appearance of a superconducting proximity effect through the LMO layer.

Fig.~\ref{fig:fig4}~(b) shows the $T$-evolution of the $I$-$V$ curves for junction $\#4$ at $H=0$. It is seen that from 200 K down to
$T_c\simeq 86$ K the resistance (voltage at a given current)
increases with decreasing $T$. This leads to qualitatively similar
insulating behavior at all bias levels at $T_{c}<T<T_{\mathrm{Curie}}$, as
shown in Figs.~\ref{fig:fig2}~(c) and \ref{fig:fig4}~(a). This trend is reversed at $T<T_c$ where the high-bias resistance
starts to decrease with decreasing $T$, compare the $I$-$V$ curves at
$T=100$, 86, 58, 30 and 1.92 K in Fig.~\ref{fig:fig4}~(b), while
the low-bias resistance continues to increase with decreasing $T$,
as seen from $R_0(T)$ in Fig.~\ref{fig:fig2}~(c) and the $dI/dV$
curves in Fig.~\ref{fig:fig4}~(e). From Fig.~\ref{fig:fig4}~(b) it
is seen that the drop in $R_{\mathrm{dc}}$ is associated with development
of an excess current in the $I$-$V$ curves at bias $V\gtrsim 0.3$ V.
Initially the excess current increases monotonously with
increasing $V$, but at high bias $V\sim 1$ V it jumps abruptly,
leading to the appearance of a step in the $I$-$V$ curves. The observed
high-bias anomaly resembles the sum-gap step in $I$-$V$'s of SIS
junctions \cite{MR_PRB2011}, but occurs at more than an order of
magnitude larger voltage. Therefore, it is not related to the
singularity in the quasiparticle DOS at the superconducting energy gap
of the YBCO electrodes \cite{MR_PRB2011,Fischer1995}. The observation and
interpretation of this high-bias anomaly at $T<T_c$ in these S/FI/S
junctions is the central new result of this work.

\begin{figure*}[t]
    \includegraphics[width=\textwidth]{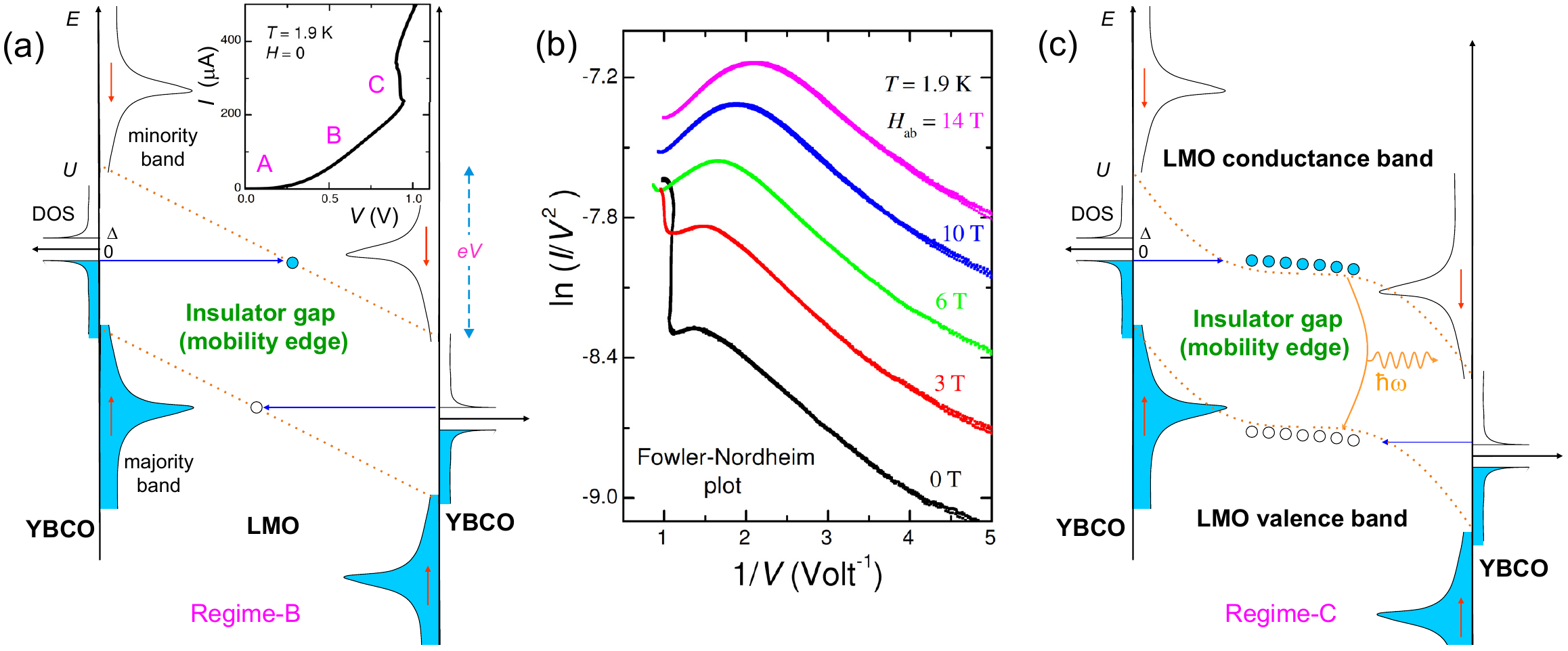}
    \caption{(Color online). (a) Sketch of the tunneling diagram of our S/FI/S junctions.
Horizontal arrows indicate Fowler-Nordheim type tunneling of
electrons into the conduction band (minority spin $\downarrow$)
and holes into the valence band (majority spin $\uparrow$) of LMO
at $eV>U+\Delta$. Inset: $I$-$V$ curve at $T=1.9$ K showing that
the current is blocked in regime-A at $V\lesssim 0.3$ V. (b)
Fowler-Nordheim plots for the junction $\#4$ at $T=1.9$ K and
different $H$. (c) Modified tunneling diagram with the proximity
effect in the conduction and valence bands of LMO. The proximity
effect leads to a reduction of the electric field in the central
part of the LMO barrier and concentration near the interfaces. The
shrinkage of the effective barrier width give rise to an excess
current.}
    \label{fig:fig5}
\end{figure*}

Fig.~\ref{fig:fig4}~(c) shows the magnetic field dependence of the
$I$-$V$ curves at $T=30$ K. It is seen that the step is suppressed
by the modest magnetic field, much smaller than the in-plane
$H_{c2}$ of YBCO. Such behavior is typical for a Josephson
current. The excess current density is in the range of kA/cm$^2$,
also typical for the Josephson current. From Fig.~\ref{fig:fig4}~(b) it is seen that the step smears out and vanishes as the
temperature increases and approaches $T_{c}$. Fig.~\ref{fig:fig4}~(d) shows the temperature dependence of the step amplitude $\Delta
I$. It follows a parabolic dependence $1-(T/T_c)^2$ (solid curve)
that is typical for the amplitude of the superconducting order
parameter. This characteristic behavior demonstrates that the
step-like increase of the excess current at large bias voltages is
a superconductivity-induced phenomenon.

Figs.~\ref{fig:fig4}~(e) and (f) show the $T$- and $H$-evolution of
the differential conductance $dI/dV$ curves (on a semi-logarithmic scale) for the same junction. At low bias voltage they reveal a thermal-activation behavior, which can be recognized
from the characteristic V-shape of the $dI/dV$ curves
\cite{Katterwe}. The well defined crossing point of the $dI/dV$
curves at $V\simeq \pm 0.3$ V in Fig.~\ref{fig:fig4}~(e) is indicative of an effective thermal
activation energy of $U\simeq 0.3$ eV \cite{Katterwe}. Notably, this is about half the value of the energy at which the polaronic band appears in the optical spectra in Figs.~\ref{fig:fig1_LMO}~(c) and \ref{fig:fig1_LMO}~(f). Therefore, we ascribe $U\simeq 0.3$ eV to the polaronic mobility
edge. At low $T$ the current through the LMO barrier is blocked at
low bias $eV<U$, as seen from the $I$-$V$ curve in the inset of Fig.~\ref{fig:fig5}~(a) (regime-A). At intermediate bias (regime-B) the
$I$-$V$ curve is non-linear at $T<T_{\mathrm{Curie}}$ and the resistance drops at
$T<T_c$, indicating the occurrence of the non-trivial long-range
superconducting proximity effect through the ferromagnetic
insulator. Finally, at high bias $V\gtrsim 1$ V (regime-C) an
excess current step appears at $T<T_c$.
%The rapid upturn of $R_0$ at $T<50$ K involves also opening of the superconducting
%gap in YBCO and possibly an additional band gap due to orbital ordering in LMO \cite{MagPolaron_NJP2004}.

\subsection{B. Fowler-Nordheim tunneling into the conduction band of the LMO}

To understand the origin of the step-like increase of the tunneling current, we sketch in Fig.~\ref{fig:fig5}~(a) the tunneling diagram of our junctions: DOS vs.
energy and the coordinate across the junction. In YBCO there are
singularities in the DOS at the superconducting gap energy of
$\Delta \sim 30$ meV \cite{MR_PRB2011,Fischer1995}. The electronic
bands in LMO are spin-polarized and are separated by the band gap
(the polaronic mobility edge) $U \sim 0.3$ eV, corresponding to
half the optical gap in Fig.~\ref{fig:fig1_LMO}~(f), followed by
maxima in the DOS at $E\sim 1$ eV
\cite{ManganitesReview,STM_FischerPRB2008,STM_YehPRB2010}. The
electric field is concentrated in the FI layer, which leads to the
linear gradient of the electric potential, that is shown by the dotted lines.

Regime-A: At low bias, $eV < U + \Delta \simeq 0.3$ V, electrons
have to tunnel through the full thickness of the LMO layer. The
probability of such a direct tunneling process is negligible due to the
large thickness of the LMO layer (20 nm). This leads to the blocking
of the tunnel current at $V \lesssim 0.3$ V that is seen from
the $dI/dV$ curves in Fig.~\ref{fig:fig4}~(e) and (f) and the
$I$-$V$ curve in the inset of Fig.~\ref{fig:fig5}~(a) (regime-A).

Regime-B: At intermediate bias, 0.3 V $<V<1$ V, a Fowler-Nordheim
type tunneling of electrons into the conduction band and holes
into the valence band of the insulator becomes possible, as
indicated by the two horizontal arrows (for electrons and holes)
in Fig.~\ref{fig:fig5}~(a). The reduction of the effective barrier
thickness leads to a rapid increase of the tunnel current at $V
\gtrsim 0.3$ V. Fig.~\ref{fig:fig5}~(b) represents the
Fowler-Nordheim plot $\ln (I/V^2)$ vs. $1/V$ for $I$-$V$'s at
different in-plane fields at $T=1.9$ K. The broad straight parts
are characteristic of the Fowler-Nordheim tunneling through the
triangular barrier \cite{FN_theory}.
% The hump %in $\ln (I/V^2)$ at small $1/V$, as marked by arrows in Fig.~3~(b), is probably
%associated with the peak in the DOS of LMO at $E=U_2$. As seen
%from the curve for $H=0$, the step-like onset of the excess
%current occurs at slightly higher voltage.
The slope of the curves depends on the barrier height $U$. It is
seen that it does not change with field, which is also seen from
the field-independence of the threshold voltages for $dI/dV$
curves in Fig.~\ref{fig:fig4}~(f).

Regime-C: At $V\gtrsim 1$ V a step in $I$-$V$ occurs. Since it
does not return back at higher bias, the excess current is present
at all bias voltages above the step. Fig.~\ref{fig:fig5}~(c)
represents our interpretation of the unusual excess-current step.
We assume that with increasing bias the Fowler-Nordheim tunneling
leads to an enhancement of the non-equilibrium population of mobile
electrons in the conduction band and holes in the valence band of LMO (beyond the polaronic mobility edge). Since the
penetration of the superconducting wave function through the now much thinner
triangular barrier is significant, such electrons acquire phase
coherence with the YBCO electrodes via the proximity effect. This
leads to a drop of the resistivity in the populated part of LMO,
which in turn leads to a distortion of the electric potential, as
shown by the dotted lines in Fig.~\ref{fig:fig5}~(c). The electric
field becomes small in the proximity-induced (central) part of the
LMO, co-aligned with the superconducting gap singularities of YBCO
electrodes. To maintain the bias voltage, the electric field has
to increase in the vicinity of the interfaces, which leads to a further shrinkage of
the tunnel barrier and to an even larger increase of the current
into the conduction band. Such a positive feedback leads to an
instability, which shows up as a step in the $I$-$V$ curve.

It should be mentioned that a similar step-like change of $I$-$V$ curves at
large bias has been reported for single manganite films and was
attributed to either depinning of polarons \cite{Rao,Wahl2003}, or
an acute Joule heating \cite{HeatJoule}. Joule heating can indeed
cause artifacts in large area films due to large currents and
dissipation powers. However, the step in $I$-$V$ curves of our
sub-micron junctions is not caused by self-heating. In general,
self-heating becomes less significant in smaller structures,
(approximately linear with size) \cite{Heating}. From Fig.~\ref{fig:fig4}~(b) it is seen that the power at the step is $\sim
0.3$ mW, which together with a realistic thermal resistance $\sim
10^4$ K/W, gives a modest heating of only several K. Most clearly,
this is seen from the connection between the step amplitude
$\Delta I$ and superconducting $T_c$ in Fig.~\ref{fig:fig4}~(d). If
heating would be significantly higher, the step would vanish at
correspondingly lower temperature with respect to $T_c$.

\begin{figure}[t]
    \includegraphics[width=2.8in]{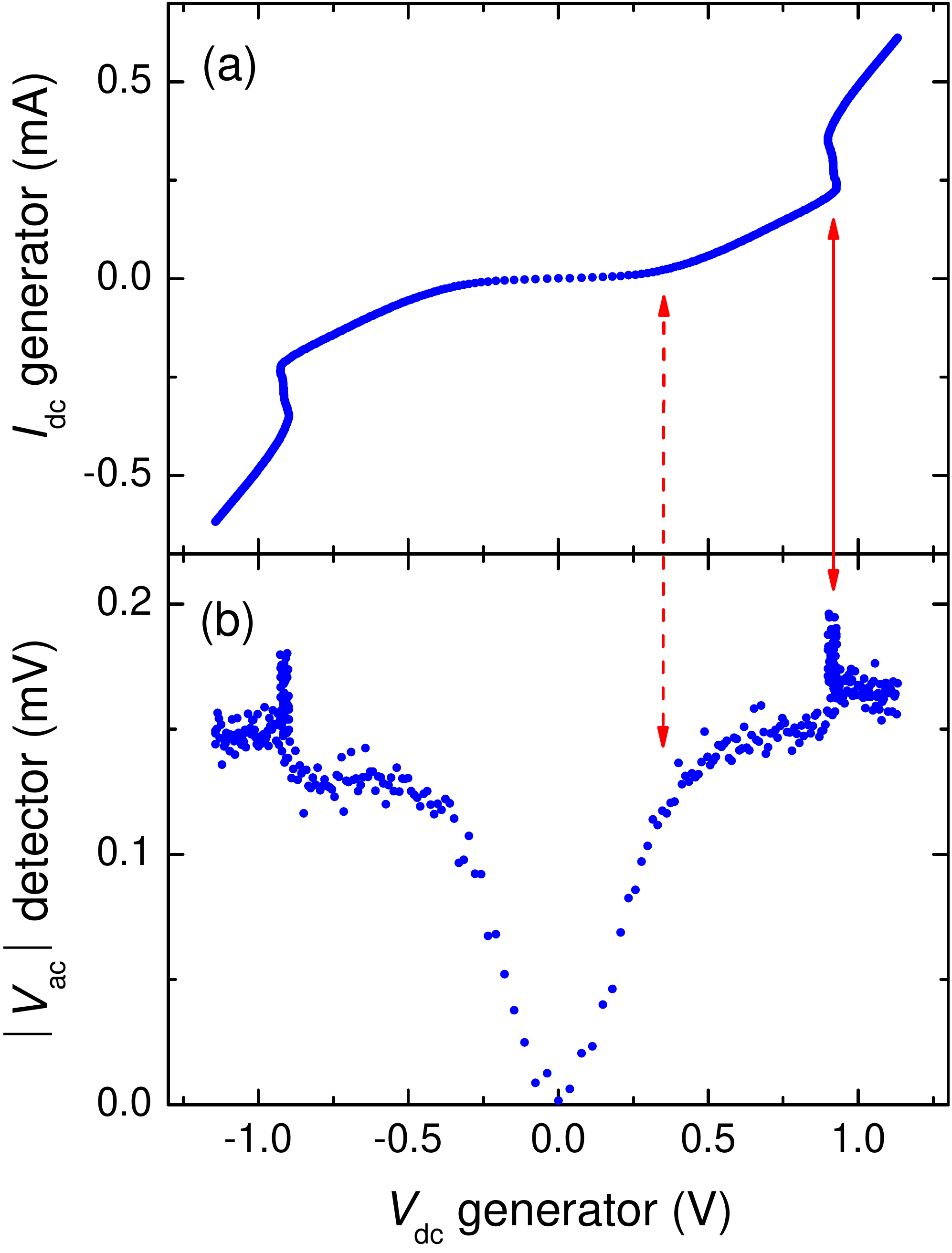}
    \caption{(Color online). Radiation detection from junction
$\#4$ at $T=1.8$ K and $H=0$. (a) The $I$-$V$ of the generator
junction $\#4$. (b) The signal of the detector junction.
Appearance of a significant photon emission above the polaronic
mobility edge in LMO (dashed line) and resonant enhancement of
emission at the step (solid line) are seen.} \label{fig:fig6}
\end{figure}

The polaron depinning mechanism of high-bias instability in LMO
\cite{Rao,Wahl2003} can be relevant, provided that by the
depinning one means that electrons become mobile if they are
lifted by the energy $U$ above $E_F$, which is equivalent to our
discussion in terms of the conduction band in LMO. This scenario
is not directly related to superconductivity. Indeed, the
reminiscence of the step at $V\sim 1$ V can be seen even at
$T=100$ K $>T_c$ in MR at $H=1$ T in Fig.~\ref{fig:fig3}~(b).
However, in our case a well defined step in the $I$-$V$'s is
observed only in the superconducting state. We thus conclude that
the new type of the superconducting proximity effect in the
conduction and valence bands of the ferromagnetic insulator
greatly amplify the phenomenon. This is also consistent with the
observed suppression of the step by a modest in-plane magnetic
field of $H=6$ T in Fig.~\ref{fig:fig4}~(c). This field is much
smaller than the in-plane $H_{c2}$ and does not significantly
affect superconductivity in the YBCO electrodes \cite{MR_PRB2011}.
However it is sufficient for suppression of the weak, proximity
induced phase coherence in the LMO layer, similar as observed for
the conventional proximity effect in S/N multilayers
\cite{NbCu1996}.

\subsection{C. Light emission at the step: evidence for the strong
non-equilibrium population in LMO}

The proposed scenario suggests the presence of a strongly
non-equilibrium electron and hole population in the conduction and
valence bands of LMO, respectively. The subsequent electron-hole
recombination should be accompanied by an inelastic emission of
bosons with an energy of $2U\sim 0.6$ eV, as indicated by the
dotted curves in Fig.~\ref{fig:fig5}~(c). The occurrence of photon
\cite{Inelastic} or phonon \cite{Polariton} emission upon
inelastic tunneling is a well known phenomenon. To check the
presence of the emission, we employed another junction on the same
chip as a radiation detector. This detector junction was
physically separated and electrically disconnected from the
generator junction to ensure that only photons, propagating
through the open space, can be detected. The generator junction
was biased with a slowly sweeping dc-voltage and a superimposed
small ac-modulation. The latter plays the role of a chopper and
modulates the radiation power at the detector. The detector
junction was biased at a constant dc-voltage of 0.5 V, slightly
above the polaronic mobility edge. The detection occurs via photon
assisted tunneling: absorbed photons create non-equilibrium hot
electrons and cause excess current through the detector junction.

Fig.~\ref{fig:fig6}~(b) shows the absolute value of the first
harmonics (chopper) signal from the detector as a function of
the dc-voltage at the generator. It is clearly seen that the emission
starts above the polaronic mobility edge in LMO at $V \gtrsim 0.3$
V (dashed arrow in Fig.~\ref{fig:fig6}). At the step a pronounced
emission enhancement takes place, as marked by the solid arrow in
Fig.~\ref{fig:fig6}. Since relaxation of non-equilibrium
quasiparticles in a superconductor is typically accompanied by
the generation of phonons, rather than photons \cite{NonEq}, the
observation of a significant photon emission supports our interpretation that inelastic
processes are taking place in the LMO layer.

\section{Conclusions}

To conclude, we have studied small all-perovskite S/FI/S tunnel
junctions made from epitaxially grown YBCO/LMO/YBCO
heterostructures by 3D-nanosculpturing. We observed a step-like
enhancement of the current through the junction at a high bias of
$V\sim 1$ V that was ascribed to a new type of superconducting
proximity effect due to non-equilibrium electrons/holes that are
injected in the conduction/valence bands of the ferromagnetic
insulator via the Fowler-Nordheim tunneling.

Since LMO is a FI with strongly spin polarized
%the FI is fully spin polarized, only the unconventional spin-tripled component of the superconducting order parameter can be sustained in its
conduction (minority) and valence (majority) bands, the observed long-range proximity effect across the 20 nm thick LMO layer, is indicative of an unconventional spin-triplet component of the superconducting order parameter. Such a long-range proximity effect through a 20 nm thick FI layer is certainly not expected for the conventional spin-singlet SC order parameter of the cuprate high $T_{c}$ superconductors.
%This conclusion is also supported by the long range (20 nm) of the proximity effect through the LMO layer, which is not expected for the spin-singlet order parameter in a ferromagnet. It should be noted, that the superconductivity in YBCO is spin-singlet.
As discussed in Refs. \cite{Efetov2005,Golubov2007}, the transformation to a so-called
odd-frequency spin-triplet order parameter requires a spin-active
interface with a non-collinear magnetic order. A modulation of the ferromagnetic order at the YBCO/manganite interface has indeed been reported in Refs. \cite{Stahn,Chakhalian,Hoppler,Kleiner,Satapathy}
%The appearance of magnetism at the YBCO/manganite interface has indeed been reported \cite{Kleiner}
and may trigger such a transformation. The
suggestion of a spin-triplet nature of the observed long-range
proximity effect in our insulating LMO is in accord with similar
reports for half-metallic manganites
\cite{Hu_LMO_Triplet2009,Dybko_LMO_Triplet2009,Kalcheim2011,Visani2012}.

The Fowler-Nordheim tunneling through the conduction/valence bands
of the FI should be fully spin-polarized. Therefore, we anticipate
that such YBCO/LMO/YBCO junctions may act as ideal spin-filter
devices for spintronic applications \cite{Moodera2011,Fert2007}.
Interestingly, spin segregation should occur in the structure. As
indicated in Fig.~\ref{fig:fig5}~(a), $\uparrow$-electrons should be
accumulated in both YBCO films (only $\downarrow$-electrons leave
the left YBCO and only $\uparrow$-electrons arrive to the right
YBCO electrode). Consequently, $\downarrow$-electrons are
accumulated in LMO.

The observation of resonant enhancement of light emission at the
step in the $I$-$V$ at $T<T_c$, Fig. \ref{fig:fig6} (b), is
consistent with similar reports in
Superconductor/Semiconductor/Superconductor light emitting diodes
\cite{Takayanagi}. The enhancement of light emission can be
attributed either to pair-wise electron-hole recombination in the
presence of Cooper pairing \cite{Takayanagi}, or to establishment
of Josephson-like phase coherence across the structure upon
emission of the photon satisfying the ac-Josephson relation, which
may even lead to lasing at high bias \cite{Nazarov}.

Finally we note that the practical applications of complex oxides is
often hindered by various materials issues. In this respect it is
important to emphasize a very good stability of our samples. No
visible deterioration was detected over a one year period at
atmospheric conditions. This indicates a very good crystalline and
chemical matching between YBCO and LMO, which
facilitates the fabrication of high-quality epitaxial heterostructures
for the use in fully spin-polarized spintronic components.

\section{Acknowledgments}
We are grateful to Ya.Fominov for valuable remarks.
Support by %the K.\:\&\:A.\:Wallenberg foundation,
the Swedish Research Council, the SU-Core facility in
Nanotechnology, the SNF Grants No. 200020-129484 and 200020-140225
and the NCCR MaNEP is gratefully acknowledged.

\end{document}